\documentclass[nohyper]{JHEP}
\usepackage{bm}
\usepackage{graphicx}

\title{Soft gluons and gauge-invariant subtractions in   NLO   
      parton-shower Monte Carlo event generators}

\author{J.C. Collins and F. Hautmann\\
        Department of Physics, Pennsylvania State University\\  
        University Park, PA 16802, USA}
      
\abstract{ We address the problem of decomposing graphs in
    perturbative QCD into terms associated with particular regions.
    Motivated by  asking how to 
    incorporate next-to-leading order (NLO) QCD corrections in
    parton-shower algorithms, we require that:  (a) The integrand for the
    hard part is to
    be integrable even if the 
    corrections are applied to a process that is  not  
    infrared and collinear safe. (b)  The splitting between the terms
    should be defined gauge-invariantly.  (c) The dependence on cut-offs
    should obey homogeneous evolution equations.
    In the context
    of  one-gluon-emission graphs for deep inelastic 
    scattering, we explain a subtractive
    technique that is based on gauge-invariant Wilson-line operators.
    Appropriate organization of subtractions involving the soft region
    allows a connection to previous work where evolution equations
    with respect to the directions of the Wilson lines have been derived.
}


\preprint{PSU-TH/230\\
September 2000\\
(Revised December 2000)}

\begin{document}

\section{Introduction}

Current parton-shower Monte Carlo event generators are essentially
leading-order (LO) QCD tools.  For precision phenomenology at present and
future high-energy colliders, it is valuable to be able to go beyond this
level of approximation~\cite{lhc99}.  Although there are a number of
treatments of various next-to-leading-order (NLO) effects in event
generators (see \cite{seymour,corcella,andre,mrenna} and references in
\cite{lhc99}), there is as yet no method for going beyond the leading
approximation systematically.  This implies that event generators are not
able to incorporate fully the known NLO (and NNLO) calculations of hard
scattering cross sections.

Recently, systematic subtractive procedures have been proposed
\cite{friberg,jccmonte,jccphi3} to correct this situation.  
An important step in the implementation of this
program is to show how to decompose Feynman graphs into sums of terms over
different regions; the terms are to be arranged to correspond to factors
in a factorization theorem that is in a form suitable for the Monte Carlo 
application.  In Ref.~\cite{jccmonte} an implementation was presented for
a simple, but phenomenologically important case: the photon-gluon fusion
process in leptoproduction. This case was simple because soft gluons do 
not enter at the leading power, so that leading regions do not overlap.  
The purpose of the present work is to show how to extend 
the method to decompose graphs with soft gluons and hence 
with overlapping leading regions.  We will treat the simplest such 
case in leptoproduction, i.e., the photon-quark process to NLO.

A Monte Carlo event
generator treats observables that are not infrared
and collinear safe.  So the kind of subtraction method used in Refs.\
\cite{IRsafeNLO} and \cite{sopnum}  
is not directly applicable, since it
relies on the calculated observable being 
infrared and collinear safe.  In particular we  
cannot use a cancellation of soft gluon contributions between
real and virtual graphs.  More general factorization theorems  
are to be used in which the soft
contributions factor instead of 
canceling \cite{back-to-back}, and 
all the factors  are defined gauge-invariantly in
terms of Wilson-line operators \cite{Sudakov}.   
We will require that the following properties be satisfied:  
\begin{itemize}
\item[(a)]  The integrand for the hard scattering coefficient 
  is to be an integrable function, and not merely a
  distribution.  Thus the hard scattering coefficient is usable even when
  one is calculating an infrared and collinear unsafe observable.  See
  Eq.\ (\ref{subtr}) for an example.  
\item[(b)] The terms in the expansion of each Feynman graph should arise
  from matrix elements of gauge-invariant operators.
\item[(c)]  In particular, 
  the necessary cut-offs on rapidity integrations are to be
  defined gauge-invariantly.  This involves the use of Wilson-line
  operators along non-lightlike directions.
\item[(d)] The evolution equations \cite{Sudakov,BalitskyBraun} with
  respect to these cut-offs should be simple, in the sense that there 
  should be no power-law remainder terms.  
  That is, the equations are homogeneous, like the renormalization-group
  equations, rather than the Callan-Symanzik equations. 
  This is achieved by the 
  technique proposed in Ref.\ \cite{nonlight}.
\end{itemize}

In Sec.~2 we  briefly describe    the framework 
provided by  the subtraction method  
for Monte Carlo event generators. In Sec.~3  
we treat  gauge-invariant subtractions and construct the 
decomposition of the partonic cross section. 
In Sec.~4 we comment on graph-by-graph subtractions. In Sec.~5 we 
give conclusions.

\section{Subtraction scheme}

To put our result in the Monte Carlo context, we 
schematically represent the
cross section in an event generator as
\begin{equation} 
\label{factorization}
   \sigma[W] = \sum_{{\rm final~states~} X} W(X)~
               {\rm{PS}} \otimes  \hat H .
\end{equation}
Here $W$ is a weight function that specifies the definition of the
particular cross section under consideration.  The symbol ${\rm{PS}}$
denotes the parton shower and the symbol $\otimes$ denotes its action on the
initial and final partons in the hard scattering, whose cross section is
denoted by $\hat H$.  In a standard Monte Carlo, the hard scattering is
taken to the leading order, $H^{\rm{(LO)}}$, and ${\rm{PS}}$ denotes
showering from the partons in $H^{\rm{(LO)}}$. In an NLO Monte Carlo, the
cross section involves a structure of the form
\begin{equation} 
\label{schemat}
   {\rm{PS}} \otimes  
   \left[ H^{\rm{(LO)}} + 
          \alpha_s  
          \left( H^{\rm{(NLO)}} 
               - {\rm{PS}}_I (1) \otimes H^{\rm(LO)} 
               - {\rm{PS}}_F (1) \otimes H^{\rm(LO)} 
          \right) 
  \right] .  
\end{equation} 
Here the first term in the square brackets is the LO hard-scattering 
function, 
and the second term is the  
subtracted NLO hard-scattering function.   
$H^{\rm(NLO)}$ is 
the result of computing the  partonic cross section from
the NLO graphs, while    
${\rm{PS}}_I (1)$ and ${\rm{PS}}_F (1)$ are the order $\alpha_s$ 
approximations to the initial-state and final-state showering. The 
subtraction terms avoid double counting of events already included 
by showering from $H^{\rm{(LO)}}$. 

We will construct a decomposition of $H^{\rm{(NLO)}}$ into 
a sum of pieces, one for each of the leading regions $R$,
\begin{equation}
\label{decomp}
H^{\rm{(NLO)}} =  \sum_{{\rm{regions}} \ R} A_H (R) \ 
+ \;\; \mbox{nonleading power} ,                   
\end{equation}
that holds uniformly over the whole of the phase 
space.  Each of the pieces in 
 (\ref{decomp}) will contain   
counterterms that prevent double counting and provide the 
suppression for going outside the region in which that particular 
piece was originally supposed to give a good approximation to the matrix 
element. 
This  is to be contrasted with 
approaches based on  splitting the 
phase space in different domains and using different approximations to 
the matrix element in these different domains (see, e.g., 
Refs.~\cite{seymour,corcella}).  

Once we have established a result of the type (\ref{decomp}), the term
corresponding to the ultraviolet region will give precisely the 
subtracted hard-scattering function to be used in
Eq.~(\ref{schemat}). The collinear terms will correspond to factors in the
cross section that are associated with showering; they therefore imply
the evolution kernels to be used in the showering.  The 
soft term would correspond to a new element in the
Monte Carlo, but,   at the order to which we work, 
we will  show that the soft term can be eliminated 
by a suitable choice of the cut-offs. This is a result analogous to
one in \cite{IRsafeNLO}.  However the results of \cite{back-to-back} 
indicate that this is unlikely to be an all-order result.

Our decomposition of graphs of order $\alpha_s$ entails a 
specific definition
of the collinear factors, which will not necessarily  coincide with
the definitions used in any current event generator.  However, we will not
address in this paper the issue of how to obtain a showering algorithm
that corresponds to our collinear terms. A correct solution of this
problem will encompass known results about coherent emission of gluons and
angular ordering of gluon emission \cite{coherent.gluons}.

\section{Decomposition of partonic cross section}

Let us consider deep inelastic scattering $\gamma^*+P \to X$, in which 
we consider a generic cross section or
observable associated with the reaction 
$\gamma^* (q) + q (p) \to g(k) + q(k^\prime)$ --- see Fig.\
\ref{fig:reaction}. 
We denote this by $\Sigma[\varphi]$, where $\varphi$ 
is a weight function that is the product of the weight function $W(X)$ 
concerned with the final state $X$ in Eq.\ (\ref{factorization}) and 
the factors in the cross section associated with the showering, including
the parton density.  Thus
$\varphi$ contains all the infrared sensitive and nonperturbative
parts of the observable.

\FIGURE{
\hspace*{3cm}
   $\raisebox{-0.5\height}{\includegraphics[scale=0.55]{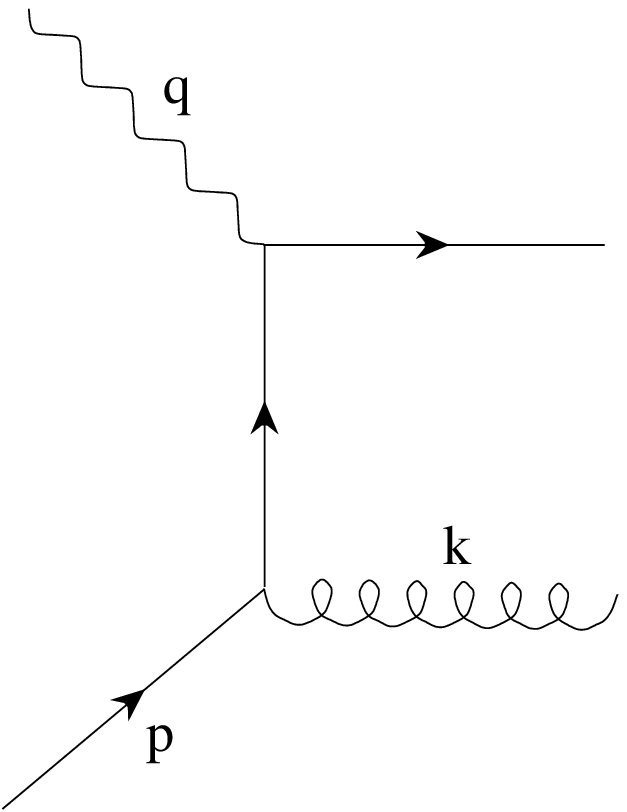}}
    ~~+~~
    \raisebox{-0.5\height}{\includegraphics[scale=0.55]{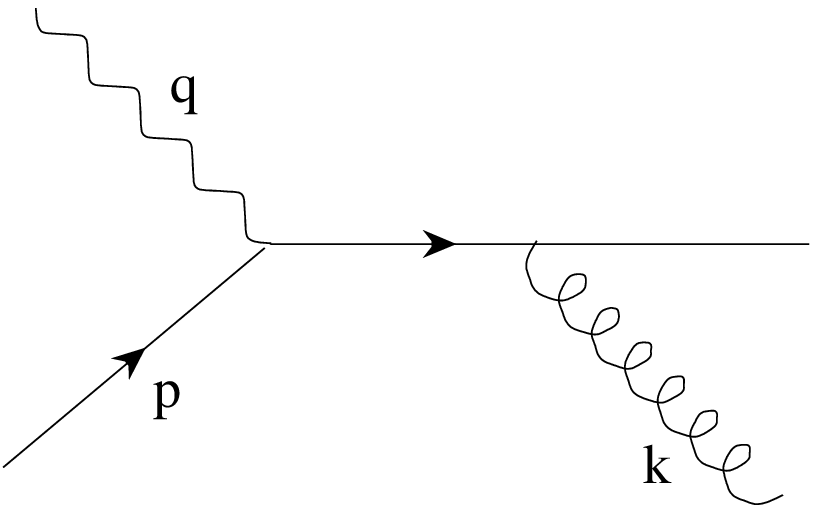}}
   $
\hspace*{3cm}
\caption{DIS with quark-induced hard scattering.}
\label{fig:reaction}}

We work in the $\gamma^*+\mbox{hadron}$ reference frame, 
and we use light-front
coordinates  $v^\mu = ( v^+, v^-, {\bf v}_T ) $ with 
$v^\pm = (v^0 \pm v^3)/\sqrt{2}$.  The hadron and photon momenta
are
$P^\mu=(P^+,m^2/2P^+,{\bf 0}_T)$ and 
$q^\mu=(-xP^+,Q^2/2xP^+,{\bf 0}_T)$. Then we parameterize the gluon
momentum $k$ as 
\begin{equation}
\label{kab}
k^\mu = \left( \alpha  x P^+, \beta { {Q^2} \over {2 x P^+}}, 
               \, \left | {\bf k}_T \right| 
               \bm{\hat \phi}
        \right) ,
\end{equation} 
where $\bm{\hat\phi}$ is a unit transverse vector at azimuthal angle
$\phi$. 

For our calculation,
the external partons are on-shell, and the incoming quark $p$ has zero
transverse momentum, so that $\Sigma$ can be written as
follows:
\begin{equation}
\label{rpiece2}
\Sigma [ \varphi] =  
\int_0^\infty d \alpha \int_0^\infty d \beta \ 
\int_0^{2 \pi} {{d \phi} \over {2 \pi}} \,  
 \varphi (x,Q^2,\alpha , \beta, \phi) \, 
 J (x,\alpha , \beta) 
\, {\cal M} (\alpha , \beta)  .
\end{equation} 
Here, $J$ is the Jacobian factor
\begin{equation}
\label{jacob}
J (x,\alpha , \beta) = { 1 \over { 16 \pi^2} } \
{ 1 \over {1 + \alpha - \beta}} \ 
\Theta \left( 
{{1-x} \over x} - \alpha \right) \ \Theta \left( 1  - 
{x \over {1-x}} \ \alpha - \beta \right)  ,          
\end{equation} 
and ${\cal M}$ is the next-to-leading-order matrix element for
$\gamma^* q$ obtained by contracting the photon Lorentz indices with
the projector corresponding to the structure function $F_2$~\cite{esw}
\begin{equation}
\label{matrel2}
{\cal M}  = 4 \ e^2_q \ g_s^2 \ C_F  \ M (\alpha , \beta) \hspace*{0.1 cm} , 
\hspace*{0.2 cm} M (\alpha , \beta) = 
(1 - \beta)^2 \ { { 1 + (1 + \alpha - \beta)^2 } 
\over { \alpha \ 
\beta \ (1 + \alpha - \beta)}} + 2 + 
6 \ {{  (1 - \beta)^2} \over { 1 + \alpha - \beta } }   .            
\end{equation} 
The physical region for $\alpha, \beta$ is the interior of the
triangle in Fig.~\ref{fig:abplane}. 

Standard arguments~\cite{libby} determine the infrared sensitive
regions contributing to the leading power behavior of $\Sigma
[\varphi]$, which are located on Fig.~\ref{fig:abplane} as follows:
The region in which the gluon is collinear to the initial
state is a neighborhood of the axis $\beta = 0$, the region in which
the gluon is collinear to the final state is a neighborhood of the
axis $\alpha = 0$, and the soft region is a neighborhood of the origin
$\alpha = 0$, $\beta = 0$.  The truly hard region lies away from the
$\alpha = 0$ and $\beta = 0$ axes.

\FIGURE{
\hspace*{3cm}
 \includegraphics[scale=0.6]{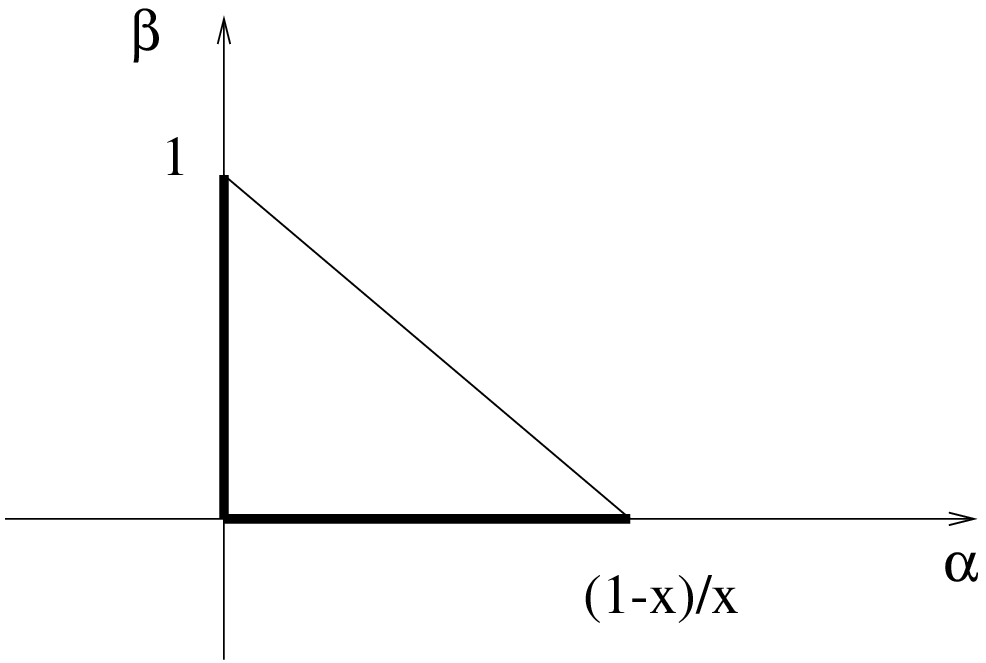}
\hspace*{3cm}
\caption{The phase space of Eq.~(\ref{rpiece2}) 
in the $\alpha, \beta$ plane. }
\label{fig:abplane}}

To obtain a
decomposition for $\Sigma $ of the type of Eq.\ (\ref{decomp}), 
we now employ the technique of Ref.~\cite{nonlight}. This    
generalizes the R-operation techniques of renormalization.  
(See  Ref.\ \cite{tka} for a related approach.)   
To ensure that the procedure is gauge-invariant, each of the terms in the
right hand side of Eq.~(\ref{decomp}) is constructed from matrix
elements involving Wilson line operators,
\begin{equation}
\label{VIVF}
V_I ( n ) = {\cal P}\exp\left(
 i g \int_{-\infty}^0 dy \, n \cdot A (y \, n ) 
 \right)
\hspace*{0.2 cm} ,  \hspace*{0.3 cm} 
V_{F} ( n ) = {\cal P}\exp\left(
  i g \int^{+\infty}_0 dy \, n \cdot A (y \, n ) 
  \right) ,  
\end{equation}
with suitable directions $n$ for the  lines. 
Evolution equations in  $n$ enable one to   
connect the results corresponding to different 
directions~\cite{Sudakov,BalitskyBraun,nonlight}. 
We define  light-like vectors ${\hat p} = \left( 1, 0, {\bf 0}_T
\right)$, ${\hat p}^\prime = \left( 0, 1, {\bf 0}_T \right)$.  We will
also use 
non-lightlike vectors $u = \left( u^+, u^-, {\bf 0}_T \right)$,
$u^\prime = \left( u^{\prime +}, u^{\prime -}, {\bf 0}_T \right)$, all of
whose components are positive.  It is
convenient to define $\eta = (2x^2 {P^+}^2/Q^2) u^- / u^+$, and  $\eta' =
(Q^2 / 2x^2 {P^+}^2) {u'}^+ / {u'}^-$.

As in \cite{nonlight}, we start with the smallest region, the soft
region $\alpha,\beta \to 0$, and determine 
the corresponding contribution to the matrix element (\ref{matrel2}):
\begin{equation} 
\label{msoft}
M_S (\alpha, \beta) =  
{2 \over { \alpha  \beta }}  - {2 \over { (\alpha + \eta^\prime 
\beta) \  \beta }} - 
{2 \over { \alpha  \ (\beta + \eta \alpha) }} .      
\end{equation}  
Observe that the first term in the right hand side of this formula is
just obtained by taking the soft approximation to Eq.~(\ref{matrel2}).
It can be thought of as the one-loop contribution to the square of a
vacuum--to--gluon matrix element of a product of eikonal Wilson lines
taken along lightlike directions ${\hat p}$, ${\hat
p}^\prime$~\cite{nonlight}.  This first term reproduces the behavior
of the matrix element $M$ when $\alpha$ and $\beta$ simultaneously
approach zero.  But there are also logarithms in its integral
associated with the collinear regions where $\alpha/\beta$ or
$\beta/\alpha$ go to zero.  The subtractions provided by the other two
terms conveniently cancel these regions.  They can be
derived from operators analogous to those for the first
term, except for replacing one of the lightlike eikonal lines by a
line along a non-lightlike direction. In particular, the second term
subtracts the divergence from a region collinear to the initial state,
i.e., from the region $\beta/\alpha \to 0$. At the same time, the
non-lightlike vector $u^\prime$ in this term provides a cut-off on the
region of small $\alpha$.  Similarly, the third term in
Eq.~(\ref{msoft}) subtracts the divergence from the region collinear
to the final state, i.e., from the region $\alpha/\beta \to 0$.  

Next we construct terms 
for the collinear regions. By applying a treatment \cite{nonlight}
analogous to that for the soft region, we arrive at  
\begin{equation} 
\label{mini}
   M_I (\alpha, \beta)
   ~=~ \frac{1}{\beta} \ 
       \frac{  1 + (1 + \alpha )^2    }
            { \alpha \ (1 + \alpha ) }
     - \frac{2}{\alpha  \beta }
     + \frac{2}{ (\alpha + \eta^\prime \beta) \  \beta }
   ~=~ \frac{1}{\beta}
       \left(
          \frac{ \alpha }{  1 + \alpha }
          + \frac{ 2 }{ \alpha + \eta^\prime \beta }
       \right)
\end{equation}
for the region collinear to the initial state, and  
\begin{eqnarray} 
\label{mfin}
   M_F (\alpha, \beta)
   &=& {1 \over \alpha} \ 
       { (1-\beta) + (1 - \beta )^3 \over \beta } \ 
       - {2 \over { \alpha  \beta }}
       + {2 \over { \alpha  \ (\beta + \eta \alpha) }}
\nonumber\\           
   &=& \frac{ 1 }{ \alpha }
       (
          -4 + 3\beta - \beta^2  
+   \frac{ 2 }{ \beta + \eta\alpha }
       )
\end{eqnarray}
for the region collinear to the final state. 
The first term of the expression in the middle 
in each of these equations is the
unsubtracted collinear approximation to the original matrix element
(i.e., the $\beta \to 0$ or $\alpha \to 0$ limit of $M$). We will   
comment below on the subtraction terms.

The fully subtracted matrix element, associated with the hard region,
is then given by
\begin{eqnarray}
\label{subtr}
M_H (\alpha , \beta) &=& M - M_S - M_I-M_F  
\nonumber\\
&=& 
(1 - \beta)^2 \ { { 1 + (1 + \alpha - \beta)^2 } 
\over { \alpha \  
\beta \  (1 + \alpha - \beta)}} + 2 + 
6 \ {{  (1 - \beta)^2} \over { 1 + \alpha - \beta } } 
\nonumber\\
&& -{1 \over \beta} \ { { 1 + (1 + \alpha )^2 } 
\over { \alpha \ 
 (1 + \alpha )}} 
- {1 \over \alpha} \ { { 1 + (1 - \beta )^2 } 
\over {  \beta }} \ (1-\beta) 
+ {2 \over { \alpha \ \beta}} 
\nonumber\\
&=& 
   \beta 
   + \frac{ \alpha }{ (1+\alpha) (1+\alpha-\beta) }
   + \frac{ 6 (1-\beta)^2 }{ (1+\alpha-\beta) }
.
\end{eqnarray} 
This matrix element is finite in all of the infrared regions. 
It can be safely integrated down to $\alpha = 0$ or  $\beta = 0$.
Moreover, it is independent of the choice of the non-lightlike directions 
$u$, $u^\prime$: the dependence on $\eta$, $\eta^\prime$ has canceled 
in Eq.~(\ref{subtr}). 

Eqs.~(\ref{msoft})--(\ref{subtr}) achieve a decomposition of the type
(\ref{decomp}) for the original matrix element.  There is one term for
each of the leading regions --- in particular, a soft term.  We now
ask: can we reorganize it in a way that is suited for a parton-shower
algorithm, 
such as, e.g., the algorithm~\cite{bengtsson} used in 
the event generators \cite{lepto}?

The soft term can be eliminated by choosing the vectors $u$ and
$u^\prime$ so that $\eta \eta^\prime = 1$; then there are only
collinear terms, as is appropriate to match the structure of the
parton-shower Monte Carlo algorithm, which has only initial-state or  
final-state branchings.  The symmetric choice $\eta=\eta^\prime = 1$
gives
\begin{equation}
\label{M1}
M_I^{({\rm{MC}})}(\alpha , \beta) =  
  \frac{ 1 }{ \beta } \ 
     \frac{ 1 + (1 + \alpha )^2 }{ \alpha \  (1 + \alpha ) }
  - \frac{ 1 }{ \alpha } \ \frac{ 2 }{ \alpha + \beta } 
,
\end{equation} 
\begin{equation}
\label{M2}
M_F^{({\rm{MC}})}(\alpha , \beta) =  
   \frac{ 1 }{ \alpha } \ 
     \frac{  1 + (1 - \beta )^2 }{ \beta }
     \ (1-\beta)  
   - \frac{ 1 }{ \beta } 
     \ \frac{ 2 }{ \alpha + \beta }
.
\end{equation}
All of the infrared contributions are now
associated with configurations that are either collinear to the
initial state or collinear to the final state.  We have inserted
superscripts in the left hand sides of Eqs.~(\ref{M1}) and (\ref{M2})
to indicate that this particular choice of the non-lightlike
directions gives rise to a structure that corresponds to that of
Monte Carlo algorithms. 

Eq.~(\ref{subtr}) determines the form of the counterterms 
to be used in the matrix element;  
Eqs.~(\ref{M1}) and (\ref{M2}) determine the form of the 
counterterms  to be used in the showering. 
The subtracted NLO matrix element (\ref{subtr}) 
only receives contributions from the truly ultraviolet 
region. As for the modification to the showering, 
 consider Eq.~(\ref{M1}). $M_I^{({\rm{MC}})}$ is associated with the 
 first branching from the initial state.  As noted below Eq.~(\ref{mfin}),  
 the first term 
in the right hand side is obtained from taking the collinear  
 approximation $\beta \to 0$ to 
the original matrix element. The coefficient  of $1/\beta$ 
is  the standard quark $\to$ quark splitting kernel:   
\begin{equation}
\label{unsubker}
 { { 1 + (1 + \alpha )^2 } 
\over { \alpha \ 
 (1 + \alpha )}} 
 = P_{qq} \left( { 1\over {1 + \alpha}} \right) 
 .                   
\end{equation}
This first term  corresponds to the standard form of the showering. 
It gives a good approximation in the initial-state collinear region,
i.e., $\beta \sim 0$.
 The second term in the right hand side of Eq.~(\ref{M1}), on the 
 other hand, is non-standard. It effectively provides a cut-off when
 $\alpha \to 0$. Note that the second term 
is suppressed in the collinear region $\beta \to 0$ at 
fixed $\alpha$. That is, the  
modified showering  coincides with the 
usual one in the collinear limit  and differs from it away from the 
collinear limit. 
Analogous remarks can be made based on the formula 
(\ref{M2}) associated with the final state.  

Observe  that if one regulated the  $\alpha \to 0$ behavior of 
the first term  in Eq.~(\ref{M1})   
by subtracting its  $\alpha \to 0$ limit, 
given by  $ 2 / ( \alpha  \beta)$, this would bring about an extra  
$\beta \to 0$ singularity. This would not be suited for our  
application in a Monte Carlo algorithm. 
In contrast, the second term in  
Eq.~(\ref{M1}) represents 
precisely what is, from our point of view, 
a  better choice of a counterterm: it subtracts the $\alpha \to 0$ 
singularity without introducing any  extra singular behavior at  
$\beta \to 0$.  

Note also that 
this counterterm   cuts off the integration over the region 
of small $\alpha$ at a value of order $\beta$: 
the leading  behavior of $M_I^{({\rm{MC}})}$ 
for small $\alpha$
is of the type 
\begin{equation}
\label{smalla}
{2 \over { \beta \ ( \alpha + \beta )}} 
+ {\mbox{regular}} \;\, {\mbox{terms}} \;\, {\mbox{in}} \;\, \alpha .
\end{equation}  
 Then we see that the procedure 
based on gauge-invariant subtractions 
that we have just applied,  
compared to the cut-off method,  
 tells us precisely  where 
the cut-off is to be placed. The position of the cut-off on $\alpha$ 
turns out to be $\beta$-dependent. In more physical terms, 
this indicates that  the cut-off to be applied in 
the initial-state shower  and the  cut-off to be applied in 
the final-state shower 
are not to be set  independently, but they are related.

\section{Graph-by-graph subtractions}

In our calculational example, the form of the four terms was determined by
requiring them to be obtainable from matrix elements of gauge-invariant
operators.  However, the procedure can be applied graph-by-graph, as we
will now explain.  The basis of this procedure is the derivation of
factorization for soft factors given in Refs.\ \cite{back-to-back,Sudakov}.

Consider the connection of a gluon to a subgraph that consists of lines
that are all
collinear in the $+$ direction.  This factor we denote by $J^\mu(k)$, where
$\mu$ is the Lorentz index that couples to the gluon.  In a region where the
gluon is either soft or collinear in the opposite direction, it is a good
approximation to replace $J^\mu$ by its $+$ component, and to replace the
momentum $k$ in $J$ by its $-$ component.  Multiplying and dividing by
$k^-$ gives
\begin{eqnarray}
  \label{eq:soft.approx}
  J^\mu(k) &\longmapsto& g^\mu_+ J^+(0,k^-,\bm{0}_T)
\nonumber\\
        &=& g^\mu_+ \frac{1}{k^-} 
            \left[ k^- J^+(0,k^-,\bm{0}_T) \right] .
\end{eqnarray}
Since the last factor is of the form of a gluon Green function contracted
with the gluon's momentum, a Ward identity can be applied.  Then after a
sum over all relevant graphs we obtain factors corresponding to matrix
elements of operators that include path-ordered exponentials
\cite{Sudakov}.

When this replacement is made, together with the corresponding replacement
for gluonic connections to the opposite ``jet'' factor, some of the terms
have new divergences when the gluon's rapidity goes to $+\infty$ or
$-\infty$, as explained earlier and in Ref.\ \cite{nonlight}.  These
divergences are cancelled by further subtractions that are constructed by
manipulations on the $1/k^-$ and $1/k^+$ factors, to give a result like
Eq.\ (\ref{msoft}).

\section{Conclusions}

In conclusion, the  above calculation  tells us 
how to organize the subtractions 
in situations in which soft gluons are present and 
both initial-state and final-state branchings contribute. This is 
one of the issues that have to be dealt with to construct 
NLO Monte Carlo event generators.  
The subtractive technique~\cite{nonlight}  used for this 
calculation is based on  graph-by-graph  R-operation methods  
and enforces gauge invariance by relating the counterterms to 
matrix elements of Wilson-line operators. 
The calculations  of this 
paper do not address the issue of how to generate the whole  shower 
corresponding to  the subtracted collinear  terms.  
The investigation of this is left to future work.

\acknowledgments

This research is supported in part by the US Department of
Energy under grant No.~DE-FG02-90ER-40577.

\end{document}